\documentclass[aps,prd,twocolumn,superscriptaddress,longbibliography]{revtex4-2}

\usepackage{amsmath,amssymb,bm}
\usepackage{graphicx}
\usepackage{physics}
\usepackage{hyperref}
\usepackage{color}

\begin{document}

\title{Static-Field Tunneling Ionization in Space-Fractional Quantum Mechanics}

\author{Marcelo F. Ciappina}
\email{marcelo.ciappina@gtiit.edu.cn}
\affiliation{Department of Physics, Guangdong Technion - Israel Institute of Technology, 241 Daxue Road, Shantou, Guangdong, 515063, China}
\affiliation{Technion - Israel Institute of Technology, Haifa, 32000, Israel}
\affiliation{Guangdong Provincial Key Laboratory of Materials and Technologies for Energy Conversion, Guangdong Technion - Israel Institute of Technology, 241 Daxue Road, Shantou, Guangdong, 515063, China}

\date{\today}

\begin{abstract}
Tunneling ionization in static or slowly varying electric fields is a cornerstone of strong-field physics and provides the entry point for semiclassical descriptions of above-threshold ionization and high-harmonic generation. In conventional quantum mechanics, the Perelomov--Popov--Terent'ev (PPT) theory and its Ammosov--Delone--Krainov (ADK) form yield an ionization rate whose defining feature is an exponential dependence governed by an under-barrier (imaginary-time) action. Here we develop an analytical ADK-like tunneling model within \emph{space-fractional} quantum mechanics, where the quadratic kinetic energy is replaced by the Riesz fractional Laplacian of order $1<\alpha\le2$. Working in a static electric field in the length gauge, we derive a closed-form tunneling exponent for a triangular exit barrier. The fractional kinetic operator deforms the conventional $I_p^{3/2}$ scaling to $I_p^{1+1/\alpha}$ and introduces a characteristic $\sin(\pi/\alpha)$ factor encoding the complex-phase structure associated with nonlocal dispersion.  We position this benchmark relative to prior tunneling studies in fractional quantum mechanics (primarily scattering through model barriers and fractal potentials) and provide a validation protocol for testing the exponent in time-dependent simulations of the fractional Schr\"odinger equation under a constant field. The result establishes a transparent reference for static-field ionization in nonlocal quantum dynamics and a baseline for strong-field approaches extensions.
\end{abstract}

\maketitle

\section{Introduction}

Strong-field ionization is a paradigmatic nonperturbative process in atomic and molecular physics. In the quasistatic regime, an external electric field distorts the binding potential, lowering and tilting the barrier such that an initially bound electronic wavefunction can escape by quantum tunneling. This mechanism underlies widely used rate models for intense-laser interactions and constitutes the first step of the semiclassical three-step picture of high-harmonic generation (ionization, propagation, and recombination). In addition to its practical value, tunneling ionization in a static field provides a uniquely stringent theoretical benchmark: it isolates barrier penetration from nonadiabatic subcycle dynamics and from recollision physics while retaining an exponentially sensitive dependence on the underlying quantum dynamics.

In conventional quantum mechanics, the theoretical description of ionization in strong laser fields originates from Keldysh theory \cite{Keldysh1965} and its subsequent refinements. In the tunneling limit, where the Keldysh parameter is small, the Perelomov--Popov--Terent'ev (PPT) theory formulates the ionization rate in terms of an imaginary-time (under-the-barrier) action accumulated during escape through the field-lowered potential barrier \cite{PPT1966,PPT1967}. The Ammosov--Delone--Krainov (ADK) model further condenses this semiclassical picture into a compact and widely used expression for atoms and ions \cite{ADK1986}. A central and robust outcome of these approaches is the exponential scaling law for the ionization rate $\Gamma \sim C_f \exp\!\left(-C_e\, I_p^{3/2}/F_0\right)$, where $F_0$ denotes the field strength and $I_p$ the ionization potential, while the prefactors $C_f$ and $C_e$ encode additional dependence on quantum numbers, angular momentum, and Coulomb corrections \cite{DeloneKrainovBook,Popov2004}. From a complementary perspective, static-field ionization can be interpreted in terms of Stark resonances and the decay of metastable states in tilted potentials, where the ionization rate is governed by barrier penetration and is exponentially small in the weak-field regime, fully consistent with the WKB/PPT/ADK logic \cite{Popov2004}. 

More broadly, strong-field phenomena such as above-threshold ionization and high-order harmonic generation are controlled by the same semiclassical action, which determines not only total ionization yields but also the phase accumulated during under-barrier motion and subsequent continuum propagation \cite{Lewenstein1994,Amini2019}. Over the past decade, refined semiclassical and numerical studies have emphasized the sensitivity of tunneling dynamics to barrier shape, dimensionality, and nonadiabatic effects, leading to systematic comparisons between simulations based on the numerical solutions of the time-dependent Schrödinger equation (TDSE), improved tunneling formulas, and Coulomb-corrected models \cite{Smirnova2007,Kaushal2013,BauerMulser1999}. This central role of the tunneling action motivates extensions of the semiclassical framework to nonstandard kinetic operators or reduced dimensionality, where modifications of the dispersion relation directly translate into altered under-barrier dynamics. In this context, fractional generalizations of the WKB action provide a natural and controlled route to assess how anomalous dispersion and nonlocal quantum transport reshape tunneling ionization, in close analogy with the strategy adopted by Tong and Lin \cite{Tong2005}, who demonstrated that accurate ionization rates across a wide range of atoms and molecules can be obtained by embedding detailed bound-state and Coulomb information into an effective semiclassical tunneling framework calibrated against TDSE simulations. In the present setting, the fractional WKB action plays a similar unifying role: it isolates the modification of the under-barrier dynamics induced by a nonquadratic dispersion relation, while allowing systematic comparison with numerical benchmarks in soft-core and reduced-dimensionality models, where the exponential dependence of the rate remains governed by the action and deviations are primarily absorbed into prefactor-like corrections.

Space-fractional quantum mechanics (SFQM) generalizes Schr\"odinger dynamics by replacing the Laplacian with a fractional (Riesz) operator, motivated by a path-integral formulation over L\'evy flights \cite{LaskinPRE2000,LaskinPRE2002}. This leads to a nonlocal kinetic term and dispersion $T(p)\propto |p|^\alpha$ with $1<\alpha\le2$ \cite{LaskinPRE2002}. The resulting nonlocality alters wave-packet spreading, modifies bound-state structure, and can produce heavy-tailed momentum distributions. Fractional Schr\"odinger dynamics has also been explored in optical analog platforms \cite{LonghiOL2015}. The formulation of SFQM and the fractional Schr\"odinger equation has been actively discussed in the literature, including clarifications regarding operator definitions and physical consistency \cite{WeiComment2016,LaskinReply2016}.

Despite a substantial amount of work on fractional Schr\"odinger dynamics, prior research on tunneling in SFQM has been mainly concerned with \emph{scattering-type} transmission problems through model barriers (e.g., delta and double-delta potentials) \cite{deOliveiraVaz2011}, as well as more recent studies of transmission through fractal (Cantor-type) potentials \cite{SinghAOP2023}. In contrast, there is not yet a standard strong-field-style benchmark for \emph{static-field ionization} in the explicit ADK/PPT sense: a closed-form tunneling exponent intended to be validated via survival-probability decay in a tilted binding potential and to serve as a baseline for time-dependent strong-field extensions. This motivates the present contribution.

The purpose of this paper is therefore to establish a transparent analytical benchmark for static-field tunneling ionization in SFQM. We focus on the tunneling \emph{exponent}, which is typically far more universal than prefactors and is the natural object for comparison between theory and numerical time propagation. We work in a static electric field (length gauge), allowing a clean separation of barrier penetration from dynamical complications.

A key modeling choice is the prefactor $D_\alpha$ multiplying the fractional Laplacian. In many contexts, $D_\alpha$ can be treated as phenomenological, effectively setting the kinetic-energy scale. Here we adopt the convention $D_\alpha=1/2$ for all $\alpha$, which preserves atomic-unit scaling, ensures a smooth $\alpha\to2$ limit, and isolates intrinsic fractional effects. With this choice, once $I_p$ is fixed (e.g., by calibrating the binding potential for each $\alpha$), differences in tunneling behavior can be attributed to the nonquadratic, nonlocal kinetic operator itself.

The aim of this contribution is thus twofold: (i) to derive a closed-form ``fractional-ADK'' (fADK) exponent that reduces exactly to conventional ADK at $\alpha=2$, and (ii) to provide a full method/validation protocol for testing the exponent in time-dependent simulations of the fractional Schr\"odinger equation under a constant field, including convergence requirements specific to nonlocal dynamics. Atomic units will be used throughout, unless otherwise stated.

\section{Conventional static-field tunneling and ADK theory}
\label{sec:ADK}

\subsection{Static-field ionization as metastable decay (Stark resonances)}
\label{subsec:stark_resonances}

A static electric field fundamentally changes the spectral character of a bound electronic state. In the absence of the field, a binding potential $V(x)$ supports stationary bound states with real eigenenergies. When a constant field $F_0$ is applied (length gauge), the potential is tilted and the bound state becomes \emph{metastable}: probability leaks to infinity through the downhill barrier, producing an outgoing wave and an exponential decay of the bound-state population. In this sense, static-field ionization can be viewed as the decay of a Stark resonance with a decay rate $\Gamma$, a perspective that connects strong-field ionization to the broader theory of field-induced metastability and tunneling decay. In the weak-field limit, the decay is controlled by barrier penetration and is exponentially small in $F_0$, consistent with the WKB approach and with the imaginary-time (saddle-point) formulations that underlie the PPT/ADK theory \cite{Keldysh1965,PPT1966,PPT1967,ADK1986,Popov2004,DeloneKrainovBook}.

\subsection{Static-field Hamiltonian and tunneling regime}
\label{subsec:static_hamiltonian}

In one dimension, the Hamiltonian for an electron bound by a potential $V(x)$ in a static electric field $F_0>0$ (length gauge) is
\begin{equation}
H = -\frac{1}{2}\frac{d^2}{dx^2} + V(x) + F_0\,x .
\label{eq:H_std}
\end{equation}
A bound state with energy $-I_p$ becomes metastable because the tilted potential admits escape on the downhill side. The ionization process is naturally separated into two regimes:

(i) the \emph{tunneling regime}, where the barrier remains above the bound-state energy and escape is dominated by under-barrier penetration, and  

(ii) the \emph{over-the-barrier} (barrier-suppression) regime, where the field lowers the saddle below $-I_p$ and the notion of tunneling ceases to control the rate.  

For Coulomb-like systems in three dimensions, a standard estimate of the barrier-suppression field gives
$F_{\rm BSI}=I_p^2/(4Z)$, where $Z$ is the charge of the ionic core \cite{DeloneKrainovBook,Popov2004}.
This estimate follows from equating the bound-state energy to the height of the Stark saddle and is frequently used as a nominal boundary between tunneling and over-the-barrier ionization.
However, as emphasized in a detailed reexamination of static and slowly ramped fields, the notion of a sharp classical or barrier-suppression threshold can be misleading in realistic situations, where adiabatic turn-on, dimensionality, and angular-momentum effects substantially shift the effective onset of ionization to larger field values \cite{CohenPRA2001}. 
In one-dimensional model potentials, an analogous threshold can be defined by the disappearance of a barrier above $-I_p$ in $V(x)+F_0\,x$, but similar caveats apply regarding its physical relevance. In the present work, we therefore focus on fields below the corrected barrier suppression values such that an exponential decay window exists and the ionization rate can be described by a semiclassical under-barrier action.
We note that the present study is formulated in one spatial dimension and is therefore intended as a benchmark model rather than a quantitatively complete description of atomic or molecular ionization. The use of a 1D geometry allows us to isolate, in the clearest and simplest possible way, how the fractional kinetic operator modifies the under-barrier dynamics, the tunneling exponent, and the survival-probability decay in a static field. In this sense, the 1D model plays the same conceptual role as in many reduced-dimensionality strong-field studies: it provides a controlled setting in which the dominant scaling of the ionization process can be identified without the additional complications of angular motion, transverse spreading, and full Coulomb geometry. At the same time, one should keep in mind that quantitatively important three-dimensional effects—such as angular-momentum structure, Coulomb focusing, and the detailed topology of the Stark saddle—are not captured in the present formulation.

\subsection{WKB momentum and under-barrier action}
\label{subsec:wkb_action}

The WKB description begins from energy conservation along the escape coordinate. For a state of energy $-I_p$, the local classical momentum satisfies
\begin{equation}
\frac{p^2(x)}{2} + V(x) + F_0\,x = -I_p .
\label{eq:energy_std}
\end{equation}
Here, it is convenient to define the barrier function
\begin{equation}
W(x)\equiv I_p + V(x) + F_0\,x .
\label{eq:W_def}
\end{equation}
Thus, the classically forbidden region is $W(x)>0$, where the momentum becomes imaginary,
\begin{equation}
p(x)= i\sqrt{2W(x)} .
\label{eq:p_wkb}
\end{equation}
The tunneling ionization probability is governed by the imaginary part of the reduced action,
\begin{equation}
S=\int_{x_i}^{x_e} p(x)\,dx,
\qquad
\Gamma\sim \exp[-2\,\mathrm{Im}\,S],
\label{eq:WKB_T}
\end{equation}
where $x_i$ and $x_e$ are the inner and outer turning points defined by $W(x)=0$.

This structure is closely connected to the imaginary-time (saddle-point) description of strong-field ionization: in PPT theory, the dominant contribution to the ionization amplitude is associated with complex-time trajectories whose action reduces, in the adiabatic/static limit, to the WKB under-barrier action \cite{PPT1966,PPT1967,Popov2004}. In ADK theory, the same exponent emerges after matching the asymptotic bound-state wavefunction to an outgoing (tunneling) solution in the field-distorted potential, yielding a compact rate formula for atoms and ions \cite{ADK1986,DeloneKrainovBook}.

\subsection{Triangular exit barrier and ADK exponent}
\label{subsec:triangular_ADK}

To obtain a simple analytical exponent that captures the universal dependence on $I_p$ and $F_0$, one focuses on the \emph{exit region} where the barrier is primarily controlled by the linear Stark term, $W(x)\simeq I_p - F_0\,x$. For short-range binding potentials, or in the far-exit region where the Coulomb tail acts as a subleading correction, it is possible to approximate
\begin{equation}
W(x)= I_p + V(x) + F_0\,x \simeq I_p - F_0\,x,
\qquad x\in[0,x_e],
\label{eq:triangular_barrier}
\end{equation}
where $x_e=\frac{I_p}{F_0}$ is the so-called exit point.
This ``triangular barrier'' approximation isolates the universal tunneling scaling and is the standard route to the ADK/PPT exponent.
More specifically, its validity is tied to the tunneling regime, where the static field is strong enough to open an exit channel but remains below the barrier-suppression threshold, so that a classically forbidden region still exists. In this case, the $x_e$ lies sufficiently far from the core, and the detailed binding potential contributes mainly as a subleading correction, allowing the barrier to be approximated by the linear Stark term. Therefore, the approximation should be understood as a benchmark for weak-to-intermediate-field tunneling ionization, rather than for the over-the-barrier regime

Substituting Eq.~\eqref{eq:triangular_barrier} into Eqs.~\eqref{eq:p_wkb}--\eqref{eq:WKB_T}, we obtain
\begin{align}
\mathrm{Im}\,S
&=
\int_0^{I_p/F_0}\sqrt{2(I_p-F_0\,x)}\,dx
\nonumber\\
&=
\frac{1}{F_0}\int_{0}^{I_p}\sqrt{2u}\,du
\qquad (u=I_p-F_0\,x)
\nonumber\\
&=
\frac{\sqrt{2}}{F_0}\,\frac{2}{3}I_p^{3/2}
=
\frac{2(2I_p)^{3/2}}{3F_0}.
\label{eq:ImS_ADK}
\end{align}
Thus, the tunneling ionization rate takes the canonical ADK/PPT exponential form \cite{ADK1986,Popov2004}
\begin{equation}
\Gamma_{\mathrm{ADK}}(F_0)\propto
\exp\!\left[-\frac{2(2I_p)^{3/2}}{3F_0}\right].
\label{eq:ADKexp}
\end{equation}

\subsection{Prefactors, Coulomb corrections, and universality}
\label{subsec:prefactor}

Equation~\eqref{eq:ADKexp} captures the most robust dependence of the ionization rate on $I_p$ and $F_0$ in the tunneling regime. In the full ADK formulation for atoms and ions, the prefactor depends on angular momentum quantum numbers and on the asymptotic form of the bound-state wavefunction, and Coulomb effects can significantly renormalize the pre-exponential factor relative to a short-range model \cite{ADK1986,DeloneKrainovBook,Popov2004}. Importantly, however, the \emph{exponent} is typically far more universal than the prefactor: it is controlled by the under-barrier action and therefore provides a sharp diagnostic for tunneling-dominated ionization.

This observation motivates the strategy adopted in the present work: we use the conventional ADK exponent as a reference and construct its deformation under a modified kinetic operator (space-fractional dispersion), focusing primarily on the exponential scaling as the most robust benchmark for comparison with numerical time propagation.

\section{Space-fractional quantum mechanics in a static electric field}
\label{sec:FQM}

\subsection{Fractional Schr\"odinger equation and Riesz operator}
\label{subsec:FSE_Riesz}

Space-fractional quantum mechanics (SFQM) is obtained by replacing the standard quadratic kinetic-energy operator in the Schr\"odinger equation with a fractional power of the Laplacian, motivated by a path-integral formulation over L\'evy flights rather than Brownian trajectories \cite{LaskinPRE2000,LaskinPRE2002}. In one spatial dimension, the time-dependent fractional Schr\"odinger equation (TDFSE) in the presence of a static electric field $F_0>0$ (length gauge) reads
\begin{equation}
i\partial_t\psi(x,t)=
\left[
D_\alpha(-\Delta)^{\alpha/2}+V(x)+F_0\,x
\right]\psi(x,t),
\quad 1<\alpha\le2.
\label{eq:FSE}
\end{equation}
Here $(-\Delta)^{\alpha/2}$ denotes the Riesz fractional Laplacian, which is most transparently defined through its action in momentum space,
\begin{equation}
\mathcal{F}\{(-\Delta)^{\alpha/2}\psi\}(k)
=
|k|^\alpha\,\tilde{\psi}(k),
\label{eq:Riesz}
\end{equation}
where $\tilde{\psi}(k)$ is the Fourier transform of $\psi(x)$ \cite{LaskinPRE2002}.

Equation~\eqref{eq:Riesz} highlights the intrinsically nonlocal character of the kinetic operator in real space: unlike the standard second derivative, the Riesz fractional Laplacian couples $\psi(x)$ to its values over the entire spatial domain. As a consequence, probability transport and boundary effects in SFQM differ qualitatively from those in conventional quantum mechanics, an issue that has been discussed extensively in the foundational literature and in subsequent clarifications \cite{WeiComment2016,LaskinReply2016}. Nevertheless, the spectral definition in Eq.~\eqref{eq:Riesz} provides a well-defined and numerically convenient starting point for both analytical considerations and time-dependent simulations.

\subsection{Dispersion relation and physical interpretation}
\label{subsec:dispersion}

The free-particle dispersion relation implied by Eq.~\eqref{eq:FSE} is
\begin{equation}
T(p)=D_\alpha |p|^\alpha .
\label{eq:dispersion}
\end{equation}
For $\alpha=2$ this reduces to the familiar quadratic dispersion $p^2/2$, while for $1<\alpha<2$ the dispersion is subquadratic and nonanalytic at $p=0$ \cite{LaskinPRE2002}. This modification has several immediate consequences: the group velocity scales as $v(p)\sim |p|^{\alpha-1}$, high-momentum components are weighted differently than in standard quantum mechanics, and wave packets exhibit anomalous spreading and heavy-tailed momentum distributions.

From the perspective of tunneling ionization, the key implication is that the relation between energy and momentum under the barrier is fundamentally altered. Since tunneling rates depend exponentially on the semiclassical action, even moderate changes in the dispersion relation can translate into significant quantitative differences in the ionization behavior. This sensitivity motivates the construction of an explicit fractional generalization of the ADK tunneling exponent, rather than relying solely on numerical results.

The prefactor $D_\alpha$ in Eq.~\eqref{eq:FSE} is instrumental in SFQM, fixing the kinetic-energy scale and, in general, may be treated as a phenomenological parameter depending on how the fractional dynamics is derived or on the effective medium under consideration \cite{LaskinPRE2000,LaskinPRE2002}. Different conventions appear in the literature, particularly in mathematical treatments where dimensional consistency is emphasized.

In the present work, we adopt the convention
\begin{equation}
D_\alpha=\frac12 \quad \text{for all }\alpha,
\label{eq:Dalpha}
\end{equation}
so that
\begin{equation}
T(p)=\frac12|p|^\alpha,
\qquad
T(p)\xrightarrow{\alpha\to2}\frac{p^2}{2}.
\label{eq:Tk}
\end{equation}
This choice preserves atomic-unit scaling, guarantees a smooth and transparent $\alpha\to2$ limit, and allows a direct deformation of conventional semiclassical tunneling theory without introducing additional $\alpha$-dependent rescalings of energy or momentum. With this convention, any change in tunneling behavior at fixed ionization potential $I_p$ can be unambiguously attributed to the nonquadratic, nonlocal kinetic operator itself.

\subsection{Static electric field and gauge considerations}
\label{subsec:gauge}

We work exclusively in the length gauge, where the static electric field enters through the linear potential term $F_0\,x$. This choice is particularly natural in the context of SFQM. In contrast to the length gauge, a velocity-gauge formulation would require a generalized minimal-coupling prescription for a fractional power of the momentum operator, which is not unique and can introduce ambiguities. By working in the length gauge, the coupling to the external field remains local in position space, and all nonlocality is confined to the kinetic operator \cite{LaskinPRE2002}.

For a static field, gauge-related subtleties associated with time-dependent vector potentials do not arise. The Hamiltonian in Eq.~\eqref{eq:FSE} therefore provides a clean and unambiguous framework for studying static-field tunneling in SFQM, directly comparable to the conventional Stark problem discussed in Sec.~\ref{sec:ADK}.

\subsection{Bound states and ionization potential in SFQM}
\label{subsec:bound_states}

Even in the absence of an external field, fractional dispersion modifies the structure of bound states. For a given binding potential $V(x)$, the ground-state energy $-I_p$ and the spatial and momentum distributions of the wavefunction generally depend on $\alpha$ \cite{LaskinPRE2002}. In particular, for $1<\alpha<2$, bound-state wavefunctions tend to exhibit heavier momentum-space tails and altered spatial localization compared to their $\alpha=2$ counterparts.

This observation motivates two complementary strategies when studying tunneling ionization in SFQM:

\emph{(i) Fixed-potential protocol:} one keeps $V(x)$ fixed across $\alpha$ and allows $I_p$ to vary. This approach captures the combined effect of fractional kinetics on bound-state structure and tunneling.

\emph{(ii) Fixed-$I_p$ protocol:} one retunes a parameter of $V(x)$ (e.g., the effective charge in a soft-core Coulomb potential) so that the ground-state ionization potential is the same for all $\alpha$. This protocol isolates the effect of the fractional kinetic operator on tunneling dynamics at a fixed binding energy.

Both strategies could be employed in numerical studies of fractional quantum systems, and their distinction is essential for interpreting ionization rates in a nonlocal kinetic theory.

\subsection{Static-field metastability and tunneling in a nonlocal theory}
\label{subsec:metastability_FQM}

When a static electric field is applied, the bound state of the fractional Hamiltonian in Eq.~\eqref{eq:FSE} becomes metastable, in direct analogy with the conventional Stark problem \cite{Keldysh1965,Popov2004}. Despite the nonlocal character of the kinetic operator, the qualitative picture of decay remains intact: probability leaks through the downhill barrier, producing an outgoing wave and an exponential decay of the bound-region population over an intermediate time window.

However, the underlying mechanism of barrier crossing is modified. Because the Riesz operator couples distant spatial points, tunneling in SFQM cannot be interpreted purely as local penetration through an infinitesimal barrier slice, as in standard WKB theory. Instead, the decay reflects a genuinely nonlocal form of barrier crossing, which nevertheless admits a well-defined semiclassical phase structure in momentum space. Related nonlocal effects on tunneling and transmission have been explored in scattering settings for model barriers \cite{deOliveiraVaz2011} and for more complex fractal potentials \cite{SinghAOP2023}.

This perspective justifies focusing on the tunneling \emph{exponent} as the primary diagnostic tool. While the detailed shape of the outgoing wave and the prefactor of the decay rate may depend sensitively on nonlocal transport and boundary effects, the exponential scaling with $I_p$ and $F_0$ provides a robust bridge between fractional and conventional quantum mechanics and sets the stage for the fractional-ADK (fADK) construction developed in the next section.

\section{Fractional-ADK tunneling exponent}
\label{sec:fADK}

In this section, we derive an ADK-like tunneling exponent appropriate for SFQM in a static electric field. The strategy closely parallels the conventional WKB/PPT/ADK construction reviewed in Sec.~\ref{sec:ADK}, but with the crucial modification that the quadratic dispersion relation is replaced by a fractional one. Throughout, we focus on the tunneling \emph{exponent}, which is the most robust and physically transparent quantity for benchmarking static-field ionization.

\subsection{Generalized energy balance and semiclassical momentum}
\label{subsec:frac_momentum}

Consider a metastable state of energy $-I_p$ governed by the fractional Hamiltonian in Eq.~\eqref{eq:FSE}. In direct analogy with conventional semiclassical theory, we introduce a local energy-balance condition along the escape coordinate,
\begin{equation}
\frac12 |p(x)|^\alpha + V(x) + F_0\,x = -I_p,
\label{eq:frac_energy}
\end{equation}
where we have used the convention $D_\alpha=1/2$ [Eq.~\eqref{eq:Dalpha}]. Defining, as before, the barrier function
\begin{equation}
W(x)\equiv I_p + V(x) + F_0\,x,
\end{equation}
the classically forbidden region corresponds to $W(x)>0$.

In this region, Eq.~\eqref{eq:frac_energy} admits complex solutions for the generalized momentum $p(x)$. 
Here, the fractional dispersion relation implies
\begin{equation}
p^\alpha = -\,2W(x),
\end{equation}
so the generalized momentum is intrinsically multivalued in the complex plane. More generally, the possible branches may be written as
\begin{equation}
p_n(x) = e^{i\left(\frac{\pi+2\pi n}{\alpha}\right)}[2W(x)]^{1/\alpha}, \qquad n\in\mathbb{Z}.
\end{equation}
A physically consistent choice of branch is to set $n=0$, i.e.
\begin{equation}
p(x)= e^{i\pi/\alpha}\,[2W(x)]^{1/\alpha},
\label{eq:p_frac}
\end{equation}
which ensures that the momentum reduces smoothly to the conventional under-barrier form $p=i\sqrt{2W}$ in the limit $\alpha\to2$. The phase factor $e^{i\pi/\alpha}$ is a direct consequence of the fractional power-law dispersion and encodes the complex-phase structure associated with nonlocal kinetic energy. This choice is the natural fractional analogue of the imaginary momentum used in standard WKB tunneling. In addition, for $1<\alpha\leq2$ it yields $\mathrm{Im}\,p>0$, which ensures an exponentially decaying wavefunction under the barrier and therefore a physically acceptable tunneling solution. By contrast, branches with $n<0$ gives the complex-conjugate momentum, $\mathrm{Im}\,p<0$, corresponding to the unphysical growing solution, while the remaining branches do not connect smoothly to the standard WKB limit and are therefore not appropriate for the semiclassical tunneling problem considered here.

\subsection{Fractional reduced action and tunneling probability}
\label{subsec:frac_action}

As in conventional tunneling theory, the transmission probability is governed by the imaginary part of the reduced action accumulated under the barrier. We therefore define
\begin{equation}
S=\int_{x_i}^{x_e} p(x)\,dx,
\qquad
\Gamma_\alpha \propto \exp\!\left[-2\,\mathrm{Im}\,S\right],
\label{eq:frac_action_def}
\end{equation}
where $x_i$ and $x_e$ are the inner and outer turning points determined by $W(x)=0$.

While the nonlocal nature of the fractional kinetic operator means that this construction cannot be interpreted as a strictly local penetration process, the momentum-space formulation underlying Eq.~\eqref{eq:p_frac} provides a well-defined semiclassical phase whose imaginary part controls the exponential decay. This approach is therefore consistent with viewing fractional tunneling as a deformation of conventional WKB theory rather than as a completely distinct mechanism.

\subsection{Triangular exit barrier approximation}
\label{subsec:frac_triangular}

To obtain a closed-form analytical expression, we now adopt the same exit-region approximation used in ADK theory. In the region where the electron exits the barrier, the linear Stark term dominates over the detailed shape of the binding potential, and we approximate
\begin{equation}
W(x)\simeq I_p - F_0\,x,
\qquad x\in[0,x_e],
\qquad x_e=\frac{I_p}{F_0}.
\label{eq:frac_triangular}
\end{equation}
This triangular-barrier approximation isolates the universal field dependence of the tunneling exponent and is justified in the same sense as in conventional ADK/PPT theory: Coulomb or short-range corrections primarily renormalize prefactors, while the leading exponential scaling is governed by the exit-region action.

Substituting Eqs.~\eqref{eq:p_frac} and \eqref{eq:frac_triangular} into Eq.~\eqref{eq:frac_action_def}, we obtain
\begin{equation}
S=\int_0^{I_p/F_0} e^{i\pi/\alpha}\,[2(I_p-F_0x)]^{1/\alpha}\,dx.
\end{equation}
Introducing the substitution $u=I_p-F_0x$ ($dx=-du/F_0$), the integral becomes
\begin{align}
S
&= \frac{e^{i\pi/\alpha}}{F_0}\,2^{1/\alpha}
\int_0^{I_p} u^{1/\alpha}\,du
\nonumber\\
&= \frac{e^{i\pi/\alpha}}{F_0}\,2^{1/\alpha}
\frac{\alpha}{\alpha+1}
I_p^{1+1/\alpha}.
\label{eq:frac_S}
\end{align}
The imaginary part of the action is therefore
\begin{equation}
\mathrm{Im}\,S=
\frac{\alpha}{\alpha+1}\,
\frac{\sin(\pi/\alpha)}{F_0}\,
2^{1/\alpha}\,
I_p^{1+1/\alpha}.
\label{eq:ImS_frac}
\end{equation}

Combining Eqs.~\eqref{eq:frac_action_def} and \eqref{eq:ImS_frac}, we obtain the central analytical result of this work: an ADK-like tunneling exponent for SFQM,
\begin{equation}
\Gamma_\alpha(F_0)\propto
\exp\!\left[
-\frac{2\alpha}{\alpha+1}\,
\frac{\sin(\pi/\alpha)}{F_0}\,
2^{1/\alpha}\,
I_p^{1+1/\alpha}
\right],
\quad 1<\alpha\le2.
\label{eq:fADK}
\end{equation}
This expression reduces exactly to the conventional ADK exponent in the limit $\alpha\to2$, since $\sin(\pi/\alpha)\to1$ and $2^{1/\alpha}\to\sqrt{2}$, yielding
\begin{equation}
\Gamma_2(F_0)\propto
\exp\!\left[-\frac{2(2I_p)^{3/2}}{3F_0}\right],
\end{equation}
in agreement with Eq.~\eqref{eq:ADKexp}.

\subsection{Physical interpretation and scope of validity}
\label{subsec:frac_interpretation}

Equation~\eqref{eq:fADK} highlights several key features of static-field tunneling in a nonlocal kinetic theory. First, the familiar $I_p^{3/2}$ scaling of the ADK exponent is replaced by $I_p^{1+1/\alpha}$, reflecting directly the fractional dispersion relation $T(p)\propto |p|^\alpha$. Second, the appearance of the factor $\sin(\pi/\alpha)$ encodes the complex-phase structure associated with the fractional power of the momentum and has no analogue in standard WKB tunneling. Third, the field dependence remains of the form $\exp(-\text{const}/F_0)$, indicating that tunneling remains exponentially suppressed in the weak-field limit despite the nonlocal character of the dynamics.

As in conventional ADK/PPT theory, Eq.~\eqref{eq:fADK} should be viewed as an expression for the leading exponential dependence of the ionization rate. Prefactors are expected to depend on the detailed form of the binding potential, on dimensionality, and on matching conditions between the bound and continuum regions. Coulomb-tail effects, in particular, may introduce additional corrections beyond the triangular-barrier approximation. Nevertheless, the exponent in Eq.~\eqref{eq:fADK} provides a robust and parameter-transparent benchmark for static-field ionization in SFQM and serves as the natural target for validation against time-dependent numerical simulations, as detailed in Sec.~\ref{sec:simulations}.

It is important to clarify how the fADK exponent in Eq.~\eqref{eq:fADK} relates to, and differs from, earlier studies of tunneling in SFQM. Most prior work on tunneling within SFQM has focused on \emph{scattering-type} problems: transmission and reflection of stationary states through finite barriers (rectangular, delta, double-delta, or fractal/Cantor-type potentials). In such settings, the primary observable is a transmission coefficient defined for an incoming plane wave of fixed energy, and the analysis emphasizes how nonlocal dispersion modifies barrier transparency relative to the conventional, $\alpha=2$, case.

The static-field ionization problem considered here is fundamentally different in both their physical setup and interpretation. First, there is no incoming flux from infinity: the system begins in a bound (or quasi-bound) state that becomes metastable when the field is applied. The relevant observable is therefore a \emph{decay rate} extracted from the time-dependent loss of probability in a bound region, rather than a transmission coefficient defined by asymptotic scattering states. Second, the energy of the escaping electron is not externally fixed but is tied to the bound-state energy $-I_p$ and to the field-induced barrier shape. This distinction mirrors the difference, already present in conventional quantum mechanics, between barrier transmission in scattering theory and tunneling ionization described by the PPT or ADK theory.

From this perspective, the fADK exponent in Eq.~\eqref{eq:fADK} should be viewed as the natural analogue of the ADK/PPT tunneling exponent, not as a fractional generalization of a transmission coefficient. The triangular exit-barrier approximation plays the same conceptual role as in standard ADK theory: it isolates the universal exit-region physics that governs the exponential suppression of ionization in the weak-field limit. Fractional tunneling results obtained for finite barriers or fractal potentials, while highly informative about nonlocal transport and interference effects, address a different physical question and are not expected to reduce to Eq.~\eqref{eq:fADK} even in appropriate limits.

Finally, the present formulation emphasizes continuity with conventional strong-field physics. The smooth $\alpha\to2$ limit of Eq.~\eqref{eq:fADK} ensures that the fractional theory deforms, rather than replaces, the standard ADK picture. This continuity is essential if SFQM is to be meaningfully embedded into the broader framework of strong-field and attosecond physics, where static-field tunneling exponents serve as reference points for understanding time-dependent ionization, continuum dynamics, and recollision-based phenomena.

\section{Results and discussion}
\label{sec:discussion}

\subsection{Physical content of the exponent}

Equation~\eqref{eq:fADK} implies: (i) a generalized binding-energy scaling $I_p^{1+1/\alpha}$ reflecting the fractional dispersion $|p|^\alpha$; (ii) an explicit nonlocal phase factor $\sin(\pi/\alpha)$ encoding the complex-branch structure of Eq.~\eqref{eq:p_frac}; and (iii) a continuous deformation of standard tunneling rather than a discontinuous breakdown at $\alpha<2$. 

\begin{figure}[t]
    \centering
    \includegraphics[width=0.48\textwidth]{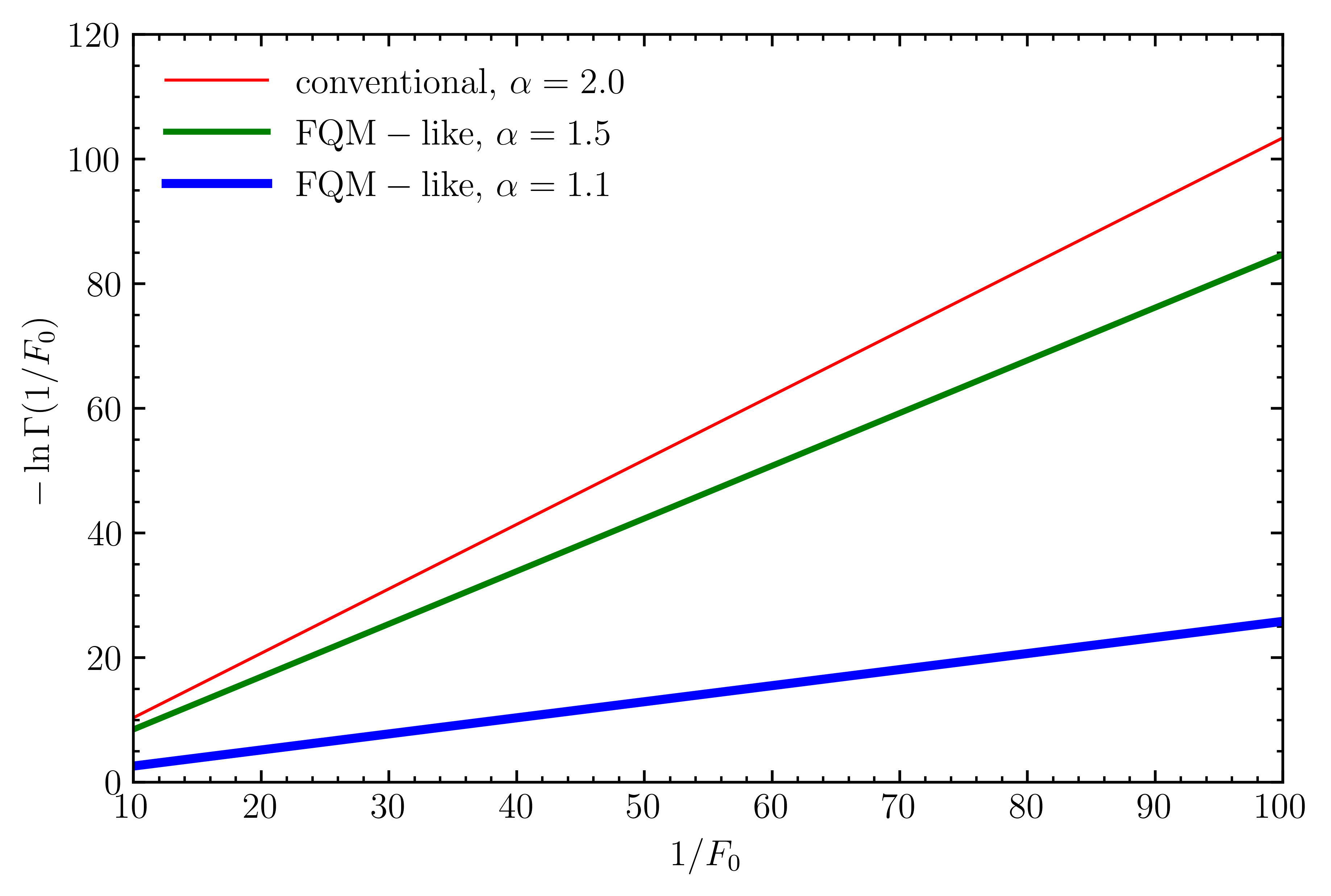}
    \caption{Comparison of $-\ln\Gamma(1/F_0)$ for the conventional ADK model ($\alpha=2.0$) and fractional-like tunneling models with $\alpha=1.5$ and $\alpha=1.1$. Here we set $I_p=-0.67$ a.u.}
    \label{fig:adk_fqm}
\end{figure}

As in conventional ADK/PPT theory, the exponent is expected to be the most robust benchmark, while prefactors depend on the core potential, matching, dimensionality, and (for Coulomb-like systems) long-range corrections \cite{ADK1986,DeloneKrainovBook,Popov2004}. In Fig.~\ref{fig:adk_fqm} we plot Eq.~\eqref{eq:fADK}, in order to compare the role of the space fractional structure on the tunneling ionization process. Specifically, we depict $-\ln\Gamma_\alpha$ versus $1/F_0$ over a field window where a clear exponential decay is observed.

From the Figure, we can discern a clear tunneling–action scaling and its systematic modification under fractional dynamics. The horizontal axis is the inverse field strength, $1/F_0$, so that moving to the right corresponds to weaker electric fields and thus a thicker/wider barrier for tunneling ionization. The vertical axis represents $-\ln\Gamma$, which in semiclassical theories is directly proportional to the under-barrier (imaginary-time) action and therefore quantifies the exponential suppression of the ionization rate: larger values of $-\ln\Gamma$ imply a much smaller $\Gamma$. The solid red curve represents the conventional case ($\alpha=2.0$), i.e.~the standard ADK-like tunneling behavior, and it increases approximately linearly with $1/F_0$, consistent with the universal scaling $\Gamma\sim\exp(-C/F_0)$. Importantly, the two SFQM-like curves (solid green for $\alpha=1.5$ and solid blue for $\alpha=1.1$) remain nearly linear as well, indicating that the process is still governed by an exponential tunneling window; however, their slopes are significantly reduced as $\alpha$ decreases. This reduction in slope signifies a smaller effective tunneling action, meaning that the exponential “penalty” for barrier penetration is weakened when fractional dispersion is introduced. Consequently, for any fixed field amplitude (fixed $1/F_0$), the fractional cases predict systematically lower values of $-\ln\Gamma$ and therefore exponentially larger ionization rates compared with the conventional ADK prediction, with the enhancement becoming especially pronounced at weak fields (large $1/F_0$). For example, in the large-$1/F_0$ region, the conventional curve implies very strong suppression, while the $\alpha=1.1$ curve yields an almost order-of-magnitude smaller tunneling exponent, corresponding to a dramatic increase in $\Gamma$. Overall, the plots highlight the main qualitative consequence of the fractional model: although ionization remains tunneling-like (exponential in $1/F_0$), the effective under-barrier action is strongly renormalized by $\alpha<2$, leading to substantially enhanced ionization yields in the sub–barrier-suppression regime.

\section{Numerical simulations: propagation, rate extraction, and validation}
\label{sec:simulations}

This section describes the numerical workflow used to validate the static-field tunneling scaling in Eq.~\eqref{eq:fADK} by direct time propagation of the fractional Schr\"odinger equation, together with a systematic extraction of decay rates and a set of convergence and robustness checks appropriate for a nonlocal kinetic operator.

We numerically solve the one-dimensional space-fractional Schr\"odinger (1D-TDFSE), Eq.~\eqref{eq:FSE}, using the Fourier symbol, Eq.~\eqref{eq:Riesz}, for the Riesz fractional Laplacian. The coordinate domain is taken as $x\in[-L,L]$ with a uniform grid of $N$ points and spacing $\Delta x=2L/N$. The corresponding wavenumber grid is defined consistently with FFT conventions,
\begin{equation}
k_n=\frac{2\pi}{2L}\times
\begin{cases}
n, & 0\le n\le N/2,\\
n-N, & N/2<n<N,
\end{cases}
\end{equation}
such that FFT-based spectral differentiation (and the multiplication by $|k|^\alpha$) is performed without additional interpolation.

To disentangle genuine fractional tunneling effects from trivial variations in binding, we employ two complementary simulation protocols.

\paragraph{Protocol A (fixed potential).}
A single binding potential $V(x)$ is chosen and kept fixed across all $\alpha$. The field-free ground-state energy $E_0(\alpha)$ then defines an $\alpha$-dependent ionization potential $I_p(\alpha)=-E_0(\alpha)$. This protocol captures how fractional kinetics modifies both the bound state (energy and spatial/momentum structure) and the subsequent ionization dynamics in the field.

\paragraph{Protocol B (fixed $I_p$).}
For each $\alpha$, the binding potential parameters are tuned such that the ionization potential matches a target value $I_p^\star$,
\begin{equation}
I_p(\alpha)=I_p^\star.
\end{equation}
In practice we use a soft-core Coulomb form,
\begin{equation}
V(x)=-\frac{Z}{\sqrt{x^2+a^2}},
\label{eq:sim_softcore}
\end{equation}
and tune $a$ at fixed $Z$ to enforce the desired $I_p^\star$. This procedure is typically used in conventional one-dimensional TDSE (1D-TDSE) simulations to mimic different atomic species.  Protocol~B isolates the influence of the \emph{kinetic operator} on tunneling at fixed binding energy and is therefore the most direct test of the $\alpha$ dependence predicted by Eq.~\eqref{eq:fADK}. The tuning of $a$ is performed by a one-dimensional root search for the function $f(a)=I_p(\alpha;a)-I_p^\star$, with bracketing and bisection.

\subsection{Ground-state preparation by imaginary-time propagation}
\label{subsec:sim_ground_state}

For each $\alpha$ (and for each set of potential parameters in Protocol~B) we compute the field-free ground state $\psi_0(x)$ of
\begin{equation}
H_\alpha(F_0=0)=\frac12(-\Delta)^{\alpha/2}+V(x)
\end{equation}
using imaginary-time propagation. Starting from a localized initial guess, we propagate
\begin{equation}
\psi(x,\tau+\Delta\tau)=\exp\!\left[-H_\alpha(F_0=0)\Delta\tau\right]\psi(x,\tau),
\label{eq:sim_imag}
\end{equation}
renormalizing the wavefunction after each step (or every few steps) to prevent collapse of the norm. A second-order split-step implementation is used,
\begin{equation}
\psi(\tau+\Delta\tau)\approx
e^{-V(x)\Delta\tau/2}\,
\mathcal{F}^{-1}\!\left[
e^{-T(k)\Delta\tau}\,\tilde{\psi}(k,\tau)
\right]\,
e^{-V(x)\Delta\tau/2},
\label{eq:sim_imag_split}
\end{equation}
which rapidly damps excited-state components. The ground-state energy is computed from the converged wavefunction via the expectation value
\begin{equation}
E_0=\langle \psi_0|H_\alpha(F_0=0)|\psi_0\rangle,
\qquad I_p=-E_0,
\end{equation}
and convergence is verified by monitoring the stabilization of $E_0$ to a prescribed tolerance.

\begin{figure}[h]
    \centering
    \includegraphics[width=0.48\textwidth]{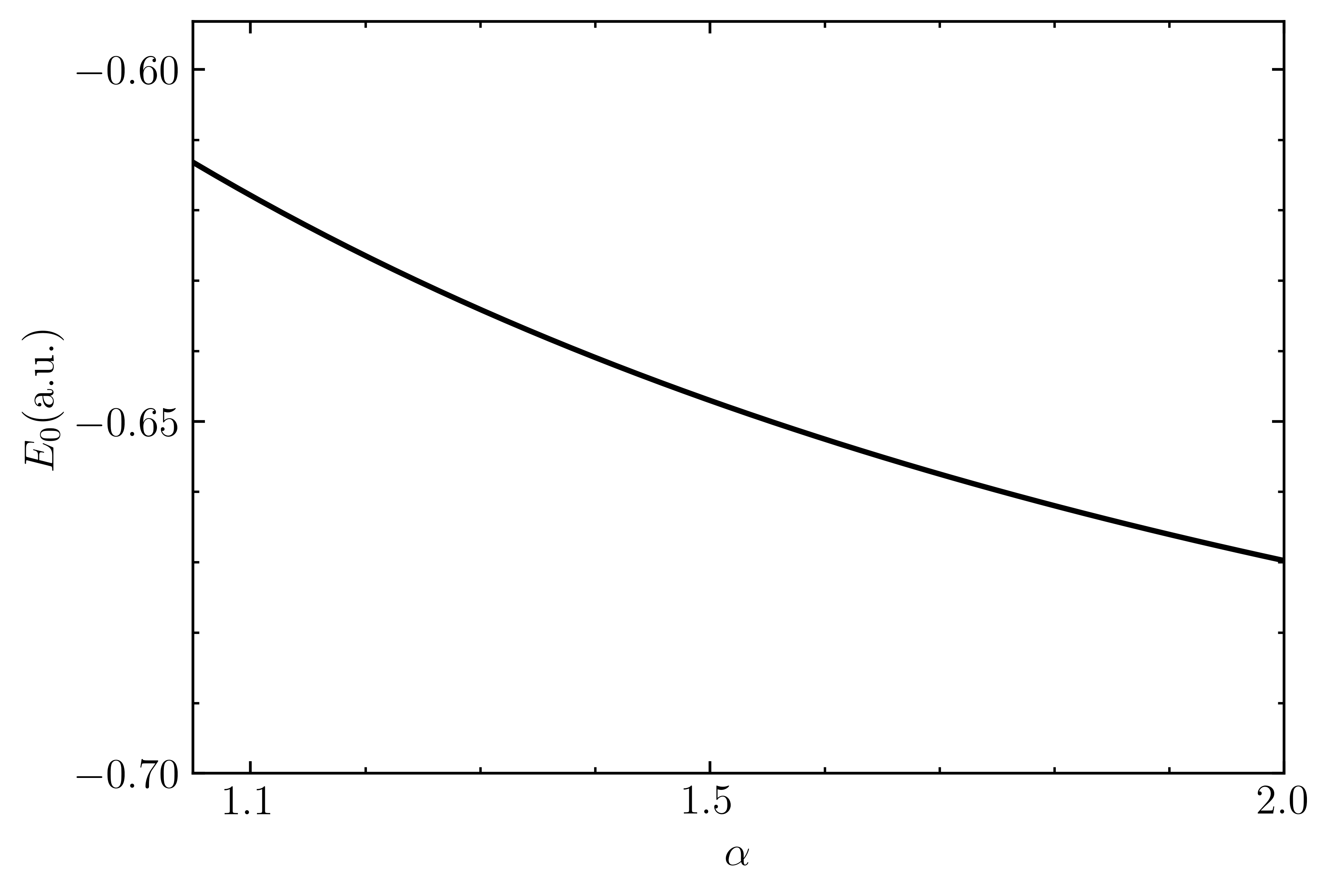}
    \caption{Ground state energy $E_0$ as a function of the fractional parameter $\alpha$, for $Z=1$ and $a=1$.}
    \label{fig:e0alpha}
\end{figure}

In Fig.~\ref{fig:e0alpha} we show the ground-state energy $E_0$ of a one-dimensional soft-core Coulomb potential, Eq.~\eqref{eq:sim_softcore}, as a function of the fractional order $\alpha$, obtained from imaginary-time propagation of the 1D-TDFSE at fixed $Z=1$ and $a=1$. A clear and systematic trend is observed: the ground-state energy becomes increasingly negative as $\alpha$ approaches the conventional value $\alpha=2$, where $E_0$ attains its minimum ($I_p=0.67$ a.u.). 

\begin{figure}[h]
    \centering
    \includegraphics[width=0.48\textwidth]{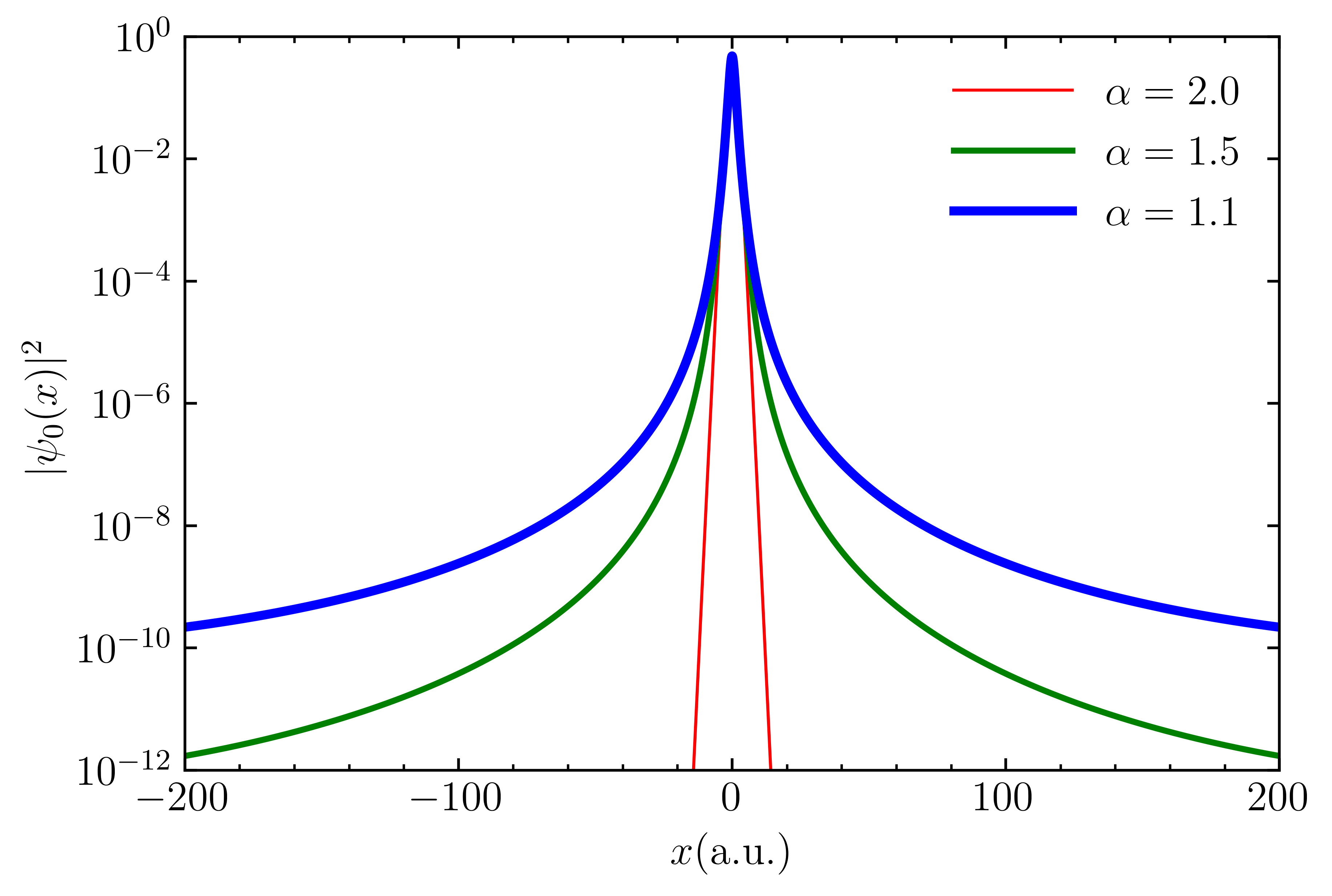}
    \caption{Ground-state probability density $|\psi_0(x)|^2$ for $\alpha=1.1$, $1.5$, and $2.0$ (semilogarithmic scale), illustrating the progressive delocalization and heavier spatial tails induced by decreasing the fractional order $\alpha$.}
    \label{fig:gslocalization}
\end{figure}

Physically, this behavior reflects the progressive restoration of the standard quadratic dispersion relation as $\alpha\to2$, which suppresses long-range, L\'evy-flight--like excursions in momentum space and enhances spatial localization of the bound state. For $\alpha<2$, the fractional kinetic operator $T(p)\propto |p|^\alpha$ penalizes small momenta less strongly than the quadratic case, effectively promoting nonlocal spreading of the wavefunction and reducing the probability density in the vicinity of the core. This enhanced delocalization weakens the binding and raises the ground-state energy. At $\alpha=2$, the balance between kinetic confinement and Coulomb attraction is optimal, yielding the deepest bound state within this family of Hamiltonians. In this sense, the minimum at $\alpha=2$ highlights that ordinary quantum mechanics corresponds to the most strongly bound realization of the Coulomb problem among its fractional generalizations, while deviations from $\alpha=2$ systematically soften the binding through the increasing nonlocality of the kinetic term. This behavior can be clearly observed in Fig.~\ref{fig:gslocalization}, where the ground-state probability density $|\psi_0(x)|^2$ reveals a clear increase of spatial delocalization as the fractional order $\alpha$ is reduced. In particular, smaller $\alpha$ leads to significantly heavier tails, while the conventional case $\alpha=2$ exhibits the fastest decay and thus the strongest localization around the ionic core.

\subsection{Real-time propagation in a constant field}
\label{subsec:sim_propagation}

At $t=0$ a constant field with peak strength $F_0$ is applied and Eq.~\eqref{eq:FSE} is propagated in real time. This static electric field is approximated by a gradual field ramp, ensuring that ionization proceeds via tunneling from a field-dressed bound state rather than being contaminated by sudden-switch-on dynamics. We use a second-order split-operator scheme,
\begin{equation}
\psi(t+\Delta t)\approx
e^{-iV_E(x)\Delta t/2}\,
\mathcal{F}^{-1}\!\left[
e^{-iT(k)\Delta t}\,\tilde{\psi}(k,t)
\right]\,
e^{-iV_E(x)\Delta t/2},
\label{eq:sim_split}
\end{equation}
with $V_E(x)=V(x)+F_0\,x$ and $T(k)$ given by Eq.~\eqref{eq:Tk}. The kinetic step is exact for the discretized spectral representation, while the overall error arises from the noncommutativity of kinetic and potential terms and scales as $\mathcal{O}(\Delta t^3)$ per step for sufficiently smooth potentials.

In practice, reliable rate extraction requires (i) a time step small enough to resolve the phase evolution in the bound region, and (ii) a total propagation time long enough to exhibit a clear exponential-decay window. The latter condition becomes more restrictive at small fields, where tunneling rates are exponentially suppressed.

Because the Riesz operator is nonlocal \cite{LaskinPRE2002}, reflections and finite-domain effects can influence the dynamics more strongly than for $\alpha=2$. We therefore combine (a) large spatial boxes with (b) smooth absorption applied only in outer buffer regions. The box size is chosen such that
\begin{equation}
L \gg x_e \sim \frac{I_p}{E},
\end{equation}
and in a way that the wavepacket remains well separated from the absorbing region over the time interval used for rate extraction.

Absorption is implemented via a multiplicative mask applied after each time step,
\begin{eqnarray}
\psi(x)&\rightarrow&M(x)\psi(x), \nonumber\\
M(x)&=&
\begin{cases}
1,& |x|\le x_{\rm cap},\\[4pt]
\exp\!\left[-\eta\left(\dfrac{|x|-x_{\rm cap}}{L-x_{\rm cap}}\right)^m\right], & |x|>x_{\rm cap},
\end{cases}
\label{eq:sim_mask}
\end{eqnarray}
with $x_{\rm cap}<L$, $\eta>0$, and $m\gtrsim 4$. The parameters are selected to minimize spurious reflections while avoiding significant absorption in the bound region. Since nonlocal coupling can, in principle, transmit boundary artifacts inward, absorber robustness is explicitly verified by varying $(x_{\rm cap},\eta,m)$ and demonstrating stable decay rates.

To quantify ionization in a manner compatible with both protocols and all $\alpha$, we define a bound-region survival probability
\begin{equation}
P_{\rm b}(t)=\int_{-x_c}^{x_c}|\psi(x,t)|^2\,dx,
\label{eq:sim_Pb}
\end{equation}
where $x_c$ encloses the ionic core and near-bound region but remains well inside the absorber onset ($x_c<x_{\rm cap}$). The choice of $x_c$ is guided by the spatial extent of the field-free ground state and is tested for stability by modest variations.

After an initial transient following the field turn-on, and before absorber/finite-box artifacts appear, $P_{\rm b}(t)$ exhibits an approximately exponential decay,
\begin{equation}
P_{\rm b}(t)\simeq P_0\,e^{-\Gamma t}.
\label{eq:sim_exp_decay}
\end{equation}
The rate $\Gamma$ is extracted by a linear regression of $\ln P_{\rm b}(t)$ over a fitting window $[t_1,t_2]$,
\begin{equation}
\ln P_{\rm b}(t)\simeq \ln P_0-\Gamma t.
\end{equation}
A diagnostic tool used to identify the exponential window is the instantaneous rate
\begin{equation}
\Gamma_{\rm inst}(t)=-\frac{d}{dt}\ln P_{\rm b}(t),
\label{eq:sim_Ginst}
\end{equation}
which displays a plateau when Eq.~\eqref{eq:sim_exp_decay} is valid. In actual calculations, a time window $[t_1,t_2]$ can be chosen by searching for the longest time interval where $\Gamma_{\rm inst}(t)$ varies only weakly around its median value; in all cases, the stability of $\Gamma$ under moderate shifts of the fitting window is verified. 

For each $\alpha$ we repeat the propagation and rate extraction for a set of static field peak strengths $\{F_{0_{j}}\}$ in the tunneling regime and obtain $\Gamma_\alpha(F_{0_{j}})$. The expected scaling is
\begin{equation}
\ln \Gamma_\alpha(F_0) \sim -\frac{C_\alpha(I_p)}{F_0}+\text{(weakly varying terms)},
\label{eq:sim_slope}
\end{equation}
where the fADK theory predicts (see Eq.~\eqref{eq:fADK})
\begin{equation}
C_\alpha(I_p)=\frac{2\alpha}{\alpha+1}\,\sin(\pi/\alpha)\,2^{1/\alpha}\,I_p^{1+1/\alpha}.
\label{eq:sim_Calpha}
\end{equation}
Accordingly, for each $\alpha$ we perform a linear fit of $\ln\Gamma_\alpha$ versus $1/F_0$ and extract the slope $-C_\alpha^{\rm (fit)}$. In Protocol~B (fixed $I_p^\star$), $C_\alpha(I_p^\star)$ becomes a direct prediction with no additional free parameters beyond the chosen convention for the kinetic prefactor, and the $\alpha$ dependence of $C_\alpha^{\rm (fit)}$ provides the most stringent test of Eq.~\eqref{eq:fADK}. In Protocol~A (fixed potential), the same analysis can be performed using the measured $I_p(\alpha)$, which tests consistency of the predicted functional dependence while retaining the physically meaningful variation of binding with $\alpha$.

\subsection{Benchmarking the 1D-TDSE}
\label{subsec:benchmark}

\begin{figure}[h]
    \centering
    \includegraphics[width=0.48\textwidth]{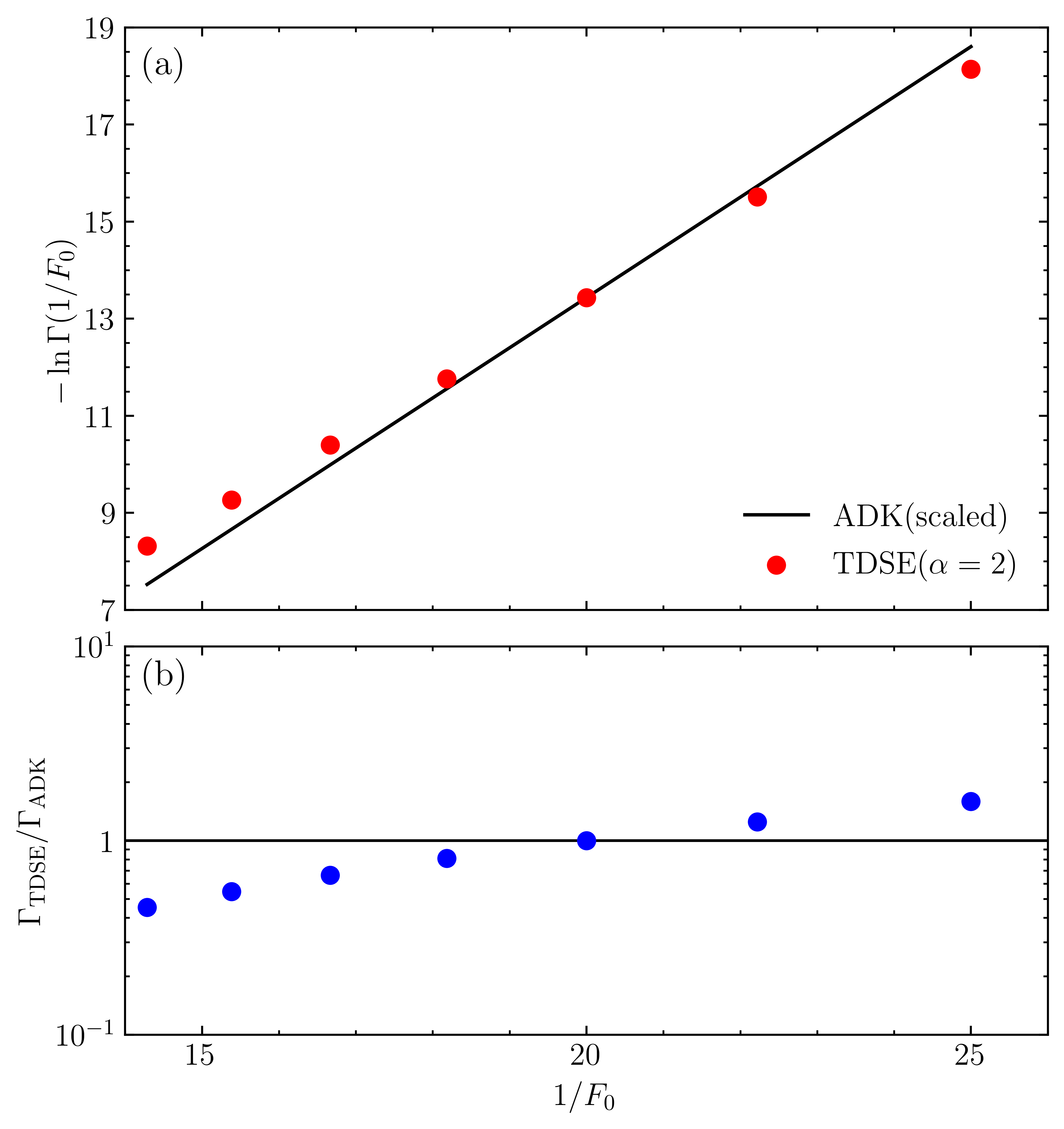}
    \caption{Benchmark of static field tunneling ionization in the standard case ($\alpha=2$).
(a) Negative logarithm of the ionization rate, $-\ln \Gamma$, as a function of the inverse field strength $1/F_0$, comparing 1D-TDSE results (symbols) with the analytical ADK tunneling prediction (solid line). (b) Ratio $\Gamma_{\mathrm{TDSE}}/\Gamma_{\mathrm{ADK}}$ shown on a logarithmic scale. The ratio remains of order unity with only weak field dependence over the field range $F_0=0.04\text{--}0.07$~a.u., demonstrating that the DC propagation and rate-extraction procedure reliably reproduce standard tunneling behavior.}
    \label{fig:benchmark}
\end{figure}

In Fig.~\ref{fig:benchmark} we present a numerical benchmark of the static field ionization protocol in the standard case $(\alpha=2)$, directly comparing 1D-TDSE results with the analytical ADK tunneling prediction. 

Figure~\ref{fig:benchmark}(a) shows the negative logarithm of the ionization rate, $-\ln\Gamma$ as a function of the inverse field strength $1/F_0$ over the range $F_0=0.04-0.07$ a.u. The TDSE data (symbols) follow an approximately linear dependence on $1/F_0$, consistent with the exponential tunneling scaling predicted by ADK theory (solid line). A single multiplicative factor is used to align the ADK curve with the numerical data at a reference field of $F_0=0.05$ a.u, reflecting the known difference between one-dimensional model calculations and the three-dimensional ADK prefactor, while preserving the field dependence.  Figure~\ref{fig:benchmark}(b) shows the ratio $\Gamma_{\mathrm{TDSE}}/\Gamma_{\mathrm{ADK}}$, on a logarithmic scale, providing a sensitive diagnostic of deviations from ADK scaling. The ratio remains of order unity with only weak field dependence across the considered range, indicating that the static propagation scheme and rate-extraction procedure reliably reproduce the expected tunneling behavior in the non-fractional case. This benchmark establishes the validity of the numerical approach and provides a controlled reference point for the fractional-kinetic calculations discussed in the following sections.

\subsection{Fractional quantum mechanics on strong field ionization}
\label{frac}

In the following section, we perform a comparative analysis of two complementary modeling protocols designed to isolate and clarify the physical consequences of SFQM on strong-field ionization. While both protocols are based on the 1D-TDFSE, they differ in how the fractional order $\alpha$ is implemented and constrained, thereby emphasizing distinct physical mechanisms.

In protocol~A, the atomic potential is kept fixed while the fractional order $\alpha$ is varied. As a consequence, the ground-state energy and ionization potential $I_p$ acquire an implicit $\alpha$ dependence. This protocol reflects a situation in which the nonlocal kinetic operator modifies the intrinsic binding properties of the system, effectively reshaping the atomic potential landscape and altering the tunneling barrier itself. Protocol~A therefore probes how fractional dynamics renormalize the bound state prior to the application of the external field.

In contrast, protocol~B is constructed to disentangle binding effects from dynamical nonlocality. Here, the ionization potential $I_p$ is held fixed across different values of $\alpha$ by tuning the soft-core parameter $a$ of the binding potential accordingly. This procedure ensures that all systems share the same static barrier height, allowing changes in the ionization rate to be attributed directly to the fractional kinetic operator rather than to trivial shifts in the ground-state energy. Protocol~B thus provides a controlled framework to assess how nonlocal propagation alone modifies the tunneling process.

By comparing these two protocols, we are able to distinguish between modifications arising from altered bound-state properties and those originating from genuinely fractional transport in the continuum. This comparison establishes a clear baseline for interpreting deviations from conventional tunneling theories and sets the stage for identifying experimentally relevant signatures of SFQM in strong-field ionization.

\begin{figure}[t]
  \centering
  \includegraphics[width=0.48\textwidth]{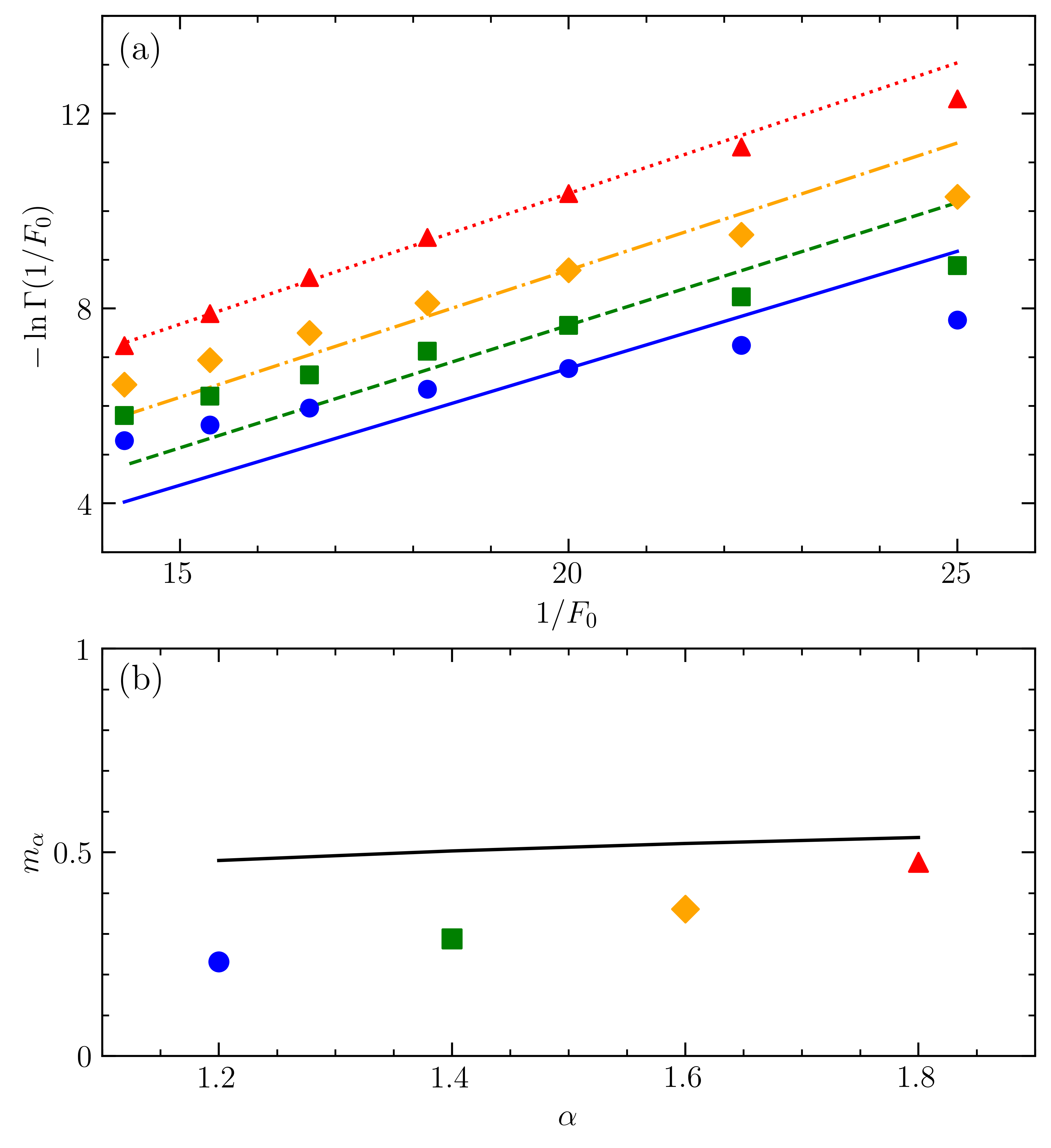}
  \caption{Comparison between numerical ionization rates obtained from the 1D-TDFSE and a fADK tunneling model for protocol~A, where the binding potential is kept fixed and the ionization potential $I_p$ varies with the fractional order $\alpha$.
(a) Action-like representation of the ionization rate, $-\ln\Gamma$, as a function of the inverse static field peak strength $1/F_0$ for $\alpha=1.2,\,1.4,\,1.6,$ and $1.8$.
Symbols denote 1D-TDFSE results (red triangle: $\alpha=1.8$, orange diamond: $\alpha=1.6$, green square: $\alpha=1.4$ and blue circles: $\alpha=1.2$), while lines correspond to the fADK scaling predictions, normalized at a reference field $F_{\mathrm{ref}}=0.05$~a.u.; different line styles are used to distinguish the various $\alpha$ values, namely dotted red: $\alpha=1.8$, dashed-dotted orange: $\alpha=1.6$, dashed green $\alpha=1.4$ and solid blue $\alpha=1.2$.
(b) Slopes $m_\alpha = d(-\ln\Gamma)/d(1/F_0)$ extracted from linear fits to panel~(a), shown as a function of the fractional order $\alpha$.
The comparison reveals a strong $\alpha$ dependence of the tunneling action originating from changes in the ground-state binding induced by the fractional kinetic operator, highlighting the role of fractional dynamics in reshaping the effective tunneling barrier.}
\label{fig:fractional_protocolA}
\end{figure}

Figure~\ref{fig:fractional_protocolA} provides a systematic comparison between numerical ionization rates obtained from the 1D-TDFSE and the corresponding fADK tunneling model. Figure~\ref{fig:fractional_protocolA}(a) shows the quantity $-\ln\Gamma$ as a function of the inverse field strength $1/F_0$ for several values of the fractional order $\alpha=1.2,\,1.4,\,1.6,$ and $1.8$, while keeping the ionization potential fixed (protocol~A). For each $\alpha$, the 1D-TDFSE results are represented by discrete symbols, whereas the fADK predictions are shown as solid lines. A key observation here is that, for all $\alpha$, the numerical 1D-TDFSE data exhibit an approximately linear dependence of $-\ln\Gamma$ on $1/F_0$ over the investigated field range, consistent with a tunneling-type exponential scaling. However, the slope of this linear behavior depends strongly on the fractional order $\alpha$. As $\alpha$ decreases below the standard quantum-mechanical value $\alpha=2$, the action extracted from the 1D-TDFSE becomes systematically smaller, indicating a substantial enhancement of the ionization rate. This trend reflects the nonlocal nature of the fractional kinetic operator, which effectively increases the probability amplitude in the classically forbidden region and facilitates barrier penetration.

The fADK curves, normalized at a reference field $F_{\mathrm{ref}}=0.05$~a.u., reproduce the overall exponential scaling and capture the monotonic dependence on $\alpha$, yet systematic deviations from the 1D-TDFSE results remain, growing rapidly as $\alpha$ is reduced. This stands in marked contrast to the conventional ($\alpha=2$) case, where 1D-TDSE simulations are accurately described by the standard ADK model. The deviations signal the limits of a simple tunneling picture when nonlocal dynamics become important. For the cited reference field, we observe that the fADK curves underestimate (overestimate) the 1D-TDSE results for $F_0<0.05$ a.u. ($F_0>0.05$ a.u.). Figure~\ref{fig:fractional_protocolA}(a) summarizes this behavior by plotting the fitted slopes $m_\alpha = d(-\ln\Gamma)/d(1/F_0)$ as a function of $\alpha$. While the fADK model predicts a smooth variation of $m_\alpha$ with fractional order, the 1D-TDFSE slopes are consistently reduced and display a stronger sensitivity to $\alpha$. Together, the two panels demonstrate that SFQM lead to a pronounced modification of tunneling ionization rates, beyond a mere rescaling of the conventional ADK exponent, highlighting the genuinely nonlocal character of strong-field ionization in the fractional regime.

\begin{figure}[t]
  \centering
  \includegraphics[width=0.48\textwidth]{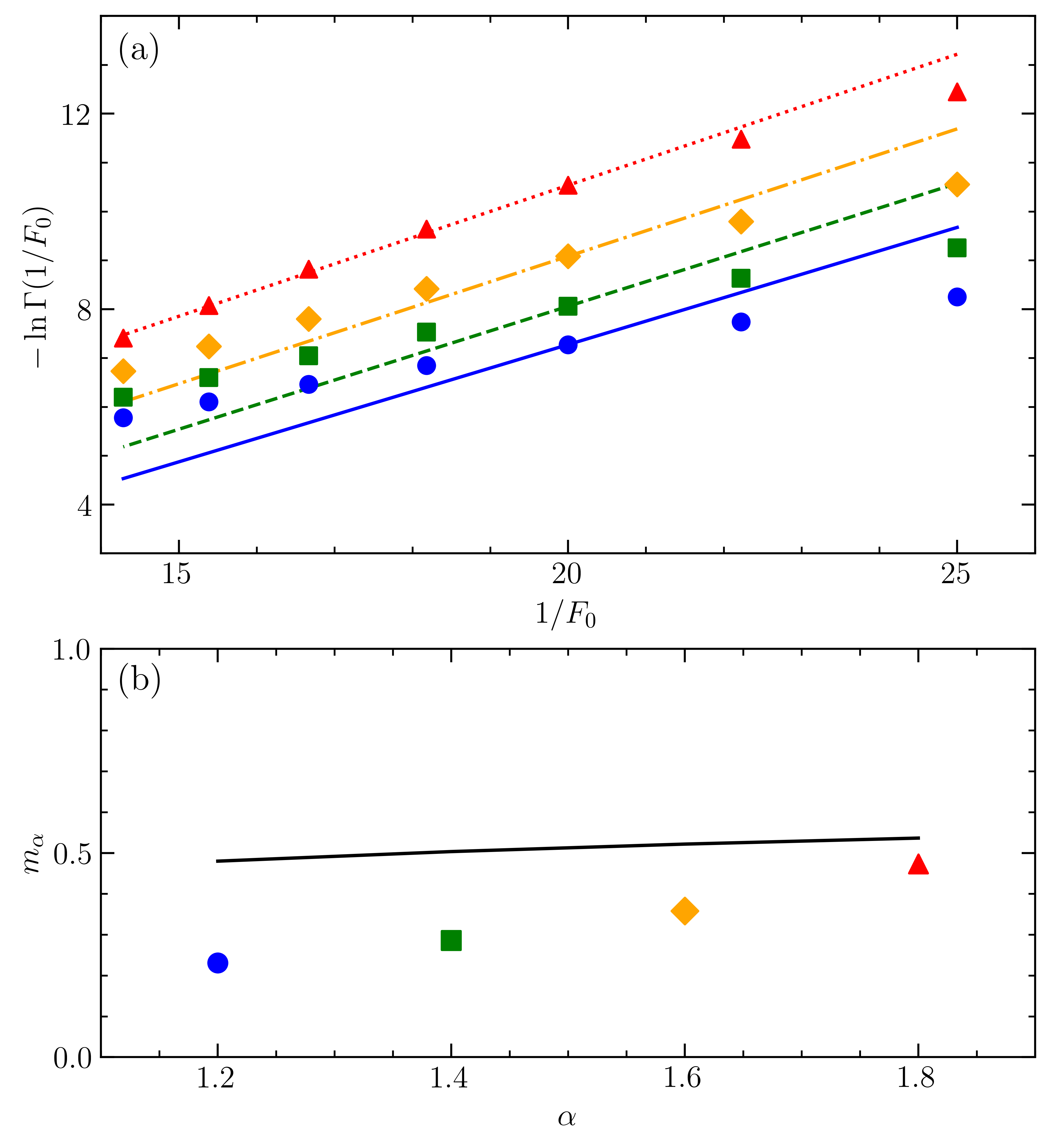}
  \caption{Fractional tunneling ionization in protocol~B, where the ionization potential $I_p$ is kept fixed across different fractional orders $\alpha$ by tuning the soft-core parameter $a$ of the binding potential.
(a) Action-like representation of the ionization rate, $-\ln\Gamma$, as a function of the inverse static field peak strength $1/F_0$ for $\alpha=1.2,\,1.4,\,1.6,$ and $1.8$.
Symbols denote 1D-TDFSE results (red triangle: $\alpha=1.8$, orange diamond: $\alpha=1.6$, green square: $\alpha=1.4$ and blue circles: $\alpha=1.2$), while lines correspond to the fADK scaling predictions, normalized at a reference field $F_{\mathrm{ref}}=0.05$~a.u.; different line styles are used to distinguish the various $\alpha$ values, namely dotted red: $\alpha=1.8$, dashed-dotted orange: $\alpha=1.6$, dashed green $\alpha=1.4$ and solid blue $\alpha=1.2$.
(b) Slopes $m_\alpha = d(-\ln\Gamma)/d(1/F_0)$ extracted from linear fits to panel~(a), plotted as a function of the fractional order $\alpha$.
Despite the identical binding energy in all cases, the 1D-TDFSE results exhibit a systematic reduction of the tunneling action with decreasing $\alpha$, demonstrating that fractional dynamics enhance tunneling through nonlocal propagation effects beyond a simple modification of the ionization potential.
}
\label{fig:fractional_protocolB}
\end{figure}

Having established in Fig.~\ref{fig:fractional_protocolA} that SFQM can strongly modify tunneling ionization when the binding properties of the system are allowed to vary with the fractional order $\alpha$ (protocol~A), we now turn to a more stringent and physically controlled test. In Fig.~\ref{fig:fractional_protocolB} we consider protocol~B, in which the ionization potential $I_p$ is held fixed for all values of $\alpha$ by appropriately tuning the soft-core parameter $a$ of the binding potential. This construction eliminates trivial changes in the barrier height and enables us to isolate the impact of the fractional kinetic operator on the tunneling dynamics itself.

Figure~\ref{fig:fractional_protocolB}(a) displays the action-like quantity $-\ln\Gamma$ as a function of the inverse static field strength $1/F_0$ for the same set of fractional orders as in Fig.~\ref{fig:fractional_protocolA}. As in the conventional case, the 1D-TDFSE results remain approximately linear in $1/F_0$, confirming that the ionization process is still governed by a tunneling-type exponential scaling. However, despite the identical binding energy in all cases, the 1D-TDFSE curves exhibit a clear and systematic dependence on $\alpha$. In particular, decreasing $\alpha$ leads to a pronounced reduction of the effective tunneling action, indicating an enhancement of the ionization rate that cannot be attributed to changes in $I_p$.

The comparison with the fADK reference curves further clarifies the origin of this effect. Although the fADK model captures the overall exponential trend and is normalized to the 1D-TDFSE data at a reference field, it does not follow the $\alpha$ dependence observed in the full 1D-TDFSE calculations. This discrepancy becomes increasingly pronounced as $\alpha$ decreases, signaling the breakdown of a purely local tunneling picture when nonlocal transport effects become significant. The slope analysis in Fig.~4(b), where $m_\alpha = d(-\ln\Gamma)/d(1/F_0)$ is plotted as a function of $\alpha$, makes this point explicit: while the fADK slopes vary only weakly with $\alpha$, the 1D-TDFSE slopes decrease substantially, providing direct evidence that fractional dynamics enhance tunneling through genuinely nonlocal propagation in the classically forbidden region.

Taken together, Figs.~\ref{fig:fractional_protocolA} and~\ref{fig:fractional_protocolB} reveal a consistent and robust trend in the strong-field ionization dynamics governed by SFQM. In both protocols, reducing the fractional order $\alpha$ leads to a systematic enhancement of the ionization rate, manifested as a reduction of the effective tunneling action. While protocol~A incorporates this behavior through $\alpha$-dependent modifications of the bound state and barrier shape, protocol~B demonstrates that a similar enhancement persists even when the ionization potential is held fixed. The close similarity between the two figures indicates that nonlocal fractional dynamics play a central role in controlling tunneling ionization, beyond simple changes in binding energy. This robustness strengthens the interpretation of the observed trends as genuine signatures of fractional quantum behavior in strong-field and tunneling-driven processes.

\section{Conclusions and outlook}

In this work, we have presented a systematic investigation of strong-field tunneling ionization within the framework of space-fractional quantum mechanics (SFQM). By solving the one-dimensional fractional time-dependent Schrödinger equation (1D-TDFSE) in static electric fields and extracting ionization rates from the decay of the bound-state population, we have established a controlled numerical platform to assess how nonlocal kinetic dynamics modify one of the most fundamental strong-field processes. Our analysis has been guided by direct comparisons with fractional generalizations of ADK-type tunneling models, allowing us to disentangle universal exponential scaling from genuinely fractional effects.

A central outcome of our study is that, across a broad range of field strengths and fractional orders, the ionization rate retains an exponential tunneling form when represented as $-\ln\Gamma$ versus $1/F_0$. This confirms that tunneling remains the dominant ionization mechanism even in the presence of fractional kinetic operators. However, the effective tunneling action is found to depend sensitively on the fractional order $\alpha$, with a systematic reduction as $\alpha$ decreases. This behavior corresponds to a pronounced enhancement of ionization relative to the conventional ($\alpha=2$) case and constitutes a robust signature of fractional quantum dynamics.

By comparing two complementary protocols, we have clarified the physical origin of this enhancement. In protocol~A, where the binding potential is kept fixed and the ionization potential varies with $\alpha$, fractional dynamics reshape both the bound state and the effective tunneling barrier, leading to reduced tunneling actions. In protocol~B, where the ionization potential is held fixed by tuning the soft-core parameter, a similar enhancement persists. The close qualitative agreement between the two protocols demonstrates that the observed trends cannot be attributed solely to changes in binding energy, but instead reflect genuinely nonlocal transport effects in the classically forbidden region. In this sense, fractional tunneling is not equivalent to conventional tunneling with a renormalized ionization potential.

Our comparison with fractional-ADK (fADK) scaling models further highlights the limits of local tunneling theories in the fractional regime. While fADK captures the overall exponential dependence on the field strength and reproduces the monotonic trend with $\alpha$, it does not follow the trend observed in the 1D-TDFSE simulations, particularly for smaller $\alpha$. This discrepancy underscores the importance of nonlocal propagation effects that are inherently beyond the scope of simple semiclassical tunneling pictures.

An important limitation of the present work is that all analytical and numerical results have been obtained within a one-dimensional model. Accordingly, the rates reported here should not be interpreted as quantitative predictions for real three-dimensional atoms or molecules. Nevertheless, we expect the main qualitative conclusions to remain relevant beyond 1D, because they originate from the modified fractional dispersion relation and the associated nonlocal transport, rather than from a specifically one-dimensional feature of the model. In particular, the persistence of an exponential tunneling window and the systematic reduction of the effective tunneling action with decreasing $\alpha$ should remain meaningful guiding signatures in higher-dimensional formulations.

Looking ahead, several extensions of the present work appear particularly promising. In particular, extending the present static-field analysis to genuine three-dimensional fractional Hamiltonians, including Coulomb and angular-momentum effects, is a natural and important next step, as it would clarify the robustness of the predicted fractional tunneling signatures in more realistic settings. More broadly, it would also be natural to generalize the present framework to time-dependent laser fields, enabling the study of fractional effects in above-threshold ionization and high-order harmonic generation. On the modeling side, the incorporation of transverse degrees of freedom, together with the development of improved semiclassical or saddle-point theories that explicitly account for fractional kinetics, remains an open and important challenge.

From a broader viewpoint, our results suggest that SFQM provides a flexible and physically transparent framework for describing enhanced tunneling and nonlocal transport phenomena in strong fields. While the fractional-ADK construction introduced here is primarily intended as a theoretical benchmark for static-field tunneling in space-fractional quantum mechanics, its broader interest lies in providing a minimal strong-field reference for nonlocal or anomalous-dispersion settings that fall outside the standard scope of conventional ADK theory. Beyond atomic ionization, similar mechanisms may play a role in condensed-matter systems, ultracold atoms, and engineered quantum platforms where effective fractional dispersion relations can be realized. We therefore anticipate that the concepts and benchmarks established here will stimulate further exploration of fractional dynamics in strong-field and tunneling-driven processes, both theoretically and experimentally.


\section*{Acknowledgments}
M.~F.~C.~acknowledges support by the National Key Research and Development Program of China (Grant No.~2023YFA1407100), Guangdong Province Science and Technology Major Project (Future functional materials under extreme conditions - 2021B0301030005), the Guangdong Natural Science Foundation (General Program project No.~2023A1515010871) and the National Natural Science Foundation of China (Grant No.~12574092). Financial support provided by the JSPS Invitational Fellowship for Research in Japan (Grant No.~S25004) is gratefully acknowledged.

\bigskip
\section*{DATA AVAILABILITY}

All numerical results presented in this article are derived from the author’s own analytical formulations and computational modeling. The data supporting these findings are available from the corresponding author upon reasonable request.

%

\end{document}